\documentclass[epj]{webofc}
\usepackage[utf8]{inputenc}
\usepackage[varg]{txfonts}   
\usepackage{booktabs}
\usepackage{xcolor}
\definecolor{darkred}{rgb}{0.4,0.0,0.0}
\definecolor{darkgreen}{rgb}{0.0,0.4,0.0}
\definecolor{darkblue}{rgb}{0.0,0.0,0.4}
\usepackage[bookmarks,linktocpage,colorlinks,
    linkcolor = darkred,
    urlcolor  = darkblue,
    citecolor = darkgreen]{hyperref}
\usepackage{subfigure}
\wocname{EPJ Web of Conferences}
\woctitle{Lattice2017}
\newcommand{\avg}[1]{\left< #1 \right>} 
\let\tilde=\widetilde
\begin{document}
%
\selectlanguage{english}
\title{%
Life Outside the Golden Window: Statistical Angles on the Signal-to-Noise Problem
}
\author{%
  \firstname{Michael} \lastname{Wagman}\inst{1,2,3}\fnsep\thanks{Speaker, gratefully acknowledges Martin Savage for contributions to the results presented. \email{mlwagman@mit.edu}\\
}
}
\institute{%
Center for Theoretical Physics, Massachusetts Institute of Technology, Cambridge, MA 02139, USA
\and
Institute for Nuclear Theory, University of Washington, Seattle, WA 98195-1550, USA
\and
Department of Physics, University of Washington, Box 351560, Seattle, WA 98195, USA
}
\abstract{%
Lattice QCD simulations of multi-baryon correlation functions can predict the structure and reactions of nuclei without encountering the baryon chemical potential sign problem. However, they suffer from a signal-to-noise problem where Monte Carlo estimates of observables have quantum fluctuations that are exponentially larger than their average values. Recent lattice QCD results demonstrate that the complex phase of baryon correlations functions relates the baryon signal-to-noise problem to a sign problem and exhibits unexpected statistical behavior resembling a heavy-tailed random walk on the unit circle. Estimators based on differences of correlation function phases evaluated at different Euclidean times are discussed that avoid the usual signal-to-noise problem, instead facing a signal-to-noise problem as the time interval associated with the phase difference is increased, and allow hadronic observables to be determined from arbitrarily large-time correlation functions.
}
\maketitle
\section{Introduction}\label{intro}


Properties of large nuclei, nuclear matter, and colliding neutron stars could be predicted from first principles if lattice quantum chromodynamics (LQCD) calculations could be performed with non-zero baryon chemical potentials.
However, the QCD action is complex in the presence of a baryon chemical potential.
Monte Carlo (MC) integration methods employing importance sampling cannot be used to compute path integrals with complex integrands, an obstacle known as the sign problem.

Nuclei can be also studied in LQCD by calculating multi-baryon correlation functions averaged over zero-density importance sampled gluon field configurations.
QCD zero-momentum nucleon correlation functions $G(t)$ have path integral and spectral representations given by
\begin{equation}
  \begin{split}
    G(t) &= \int \mathcal{D}U\; e^{-S(U)}\; C(t;U) = \sum_n \tilde{Z}_n Z_n^\dagger e^{-E_n t} \sim e^{-M_N t},
  \end{split}\label{eq:Gdef}
\end{equation}
where $t$ denotes the Euclidean time separation of the correlation function source and sink,
$U_\mu(x) \in SU(3)$ denotes the gluon field,
$S(U)$ denotes the gluon action after quarks have been integrated out,
$C(t;U)$ denotes nucleon correlation functions computed in gluon field configuration $U$,
$n$ labels QCD eigenstates, 
$E_n$ denotes their energies, 
$Z_n$ ($\tilde{Z}_n$) are overlap factors for source (sink) interpolating operators, $\sim$ denotes proportionality as $t \rightarrow \infty$, 
and $M_N$ is the nucleon mass.
Lattice units are used throughout.
The path integral over gluon field configurations can be numerically approximated using a MC ensemble of $i=1,\cdots,N$ correlation functions $C_i(t) = C(t,U_i)$ computed in zero-density importance sampled gluon field configurations, 
\begin{equation}
  \begin{split}
    G(t) = \avg{C_i(t)} = \sum_{i=1}^N C_i(t),
  \end{split}\label{eq:GNsamp}
\end{equation}
where equality holds in the limit $N\rightarrow \infty$.
An effective mass that becomes equal to the nucleon mass in the limit $t\rightarrow \infty$ can be constructed as $M(t) = - \partial_t \ln G(t) = \ln G(t+1) - \ln G(t)$.

Parisi~\cite{Parisi:1983ae} and Lepage~\cite{Lepage:1989hd} showed that the signal-to-noise (StN) ratios of correlation functions in MC calculations can be understood physically.
The variance of the real part of $C_i$ includes contributes from $C_i^2$, which has two-nucleon quantum numbers, and from $|C_i|^2$, which has nucleon-antinucleon quantum numbers.
The lowest-energy state with nucleon-antinucleon quantum numbers and no valence quark annihilation is composed of three pions,
and so the signal-to-noise (StN) ratio of nucleon correlation functions decays exponentially as
\begin{equation}
  \begin{split}
    \text{StN}(C_i)  \sim \frac{\avg{C_i}}{\sqrt{\avg{|C_i|^2}}} \sim e^{-(M_N - \frac{3}{2}m_\pi)t}.
  \end{split}\label{eq:ParisiLepage}
\end{equation}
LQCD studies have revealed a golden window of intermediate times where $C_i$ is described by ground-state time evolution but $|C_i|^2$ is not dominated by its $3\pi$ ground state~\cite{Beane:2009kya,Beane:2009gs,Beane:2009py}.
The StN ratio degrades exponentially less quickly in the golden window than at large times.
Studies of small nuclei through the golden window have shown recent success,
and it is possible to extend the golden window by reducing overlap with the variance ground state~\cite{Detmold:2014rfa,Detmold:2014hla}.
Still, the golden window shrinks with increasing baryon number, 
and
,
different analysis strategies may be required when studying large nuclei whose correlation functions have no apparent golden window.

\section{Complex Correlation Function Statistics}


Baryon correlation functions are complex in generic gluon field configurations, and their statistics can be usefully analyzed in terms of their log-magnitudes and complex phases~\cite{Wagman:2016bam},
\begin{equation}
  \begin{split}
    C_i = |C_i|e^{i\theta_i} = e^{R_i + i \theta_i}.
  \end{split}\label{eq:magphase}
\end{equation}
If the nucleon correlation function was computed directly with MC importance sampling of some $e^{-S_{eff}(U)}$, then the effective action used would be $S_{eff}(U) = S(U) - R(t;U) - i\theta(t;U)$.
Non-zero $\theta_i$ in generic gluon field configurations gives direct MC importance sampling of the nucleon correlation function a sign problem.
The standard method of calculating average correlation functions with an ensemble of zero-density field configurations can be viewed as reweighting the sign problem associated with $e^{i\theta_i}$.
Reweighting approaches to sign problems generically introduce StN problems, suggesting that the baryon correlation function phase is related to the StN problem.

\begin{figure}[t] 
  \centering
  \includegraphics[width=\columnwidth,clip]{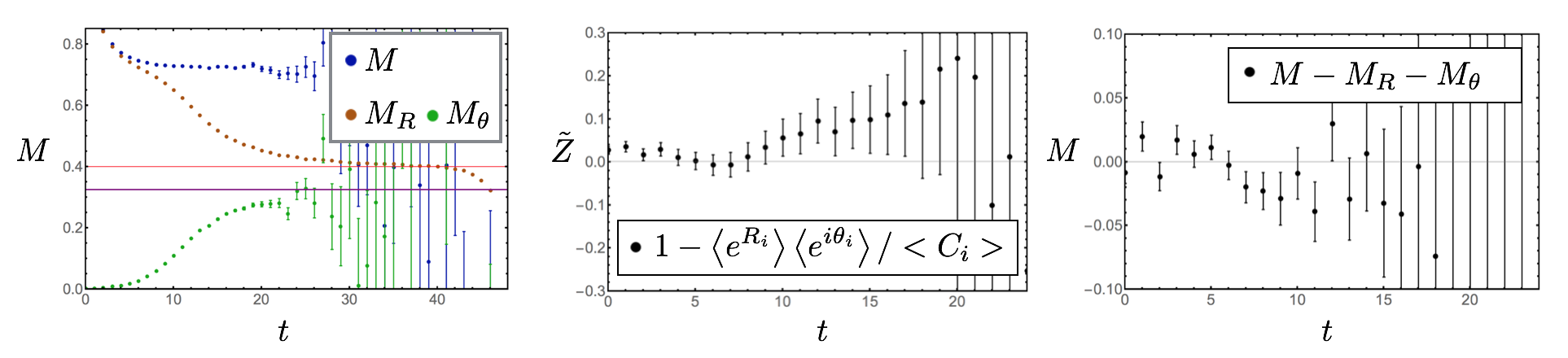}
  \caption{The left plot shows the nucleon effective mass as well as the magnitude and phase effective masses of Eq.~\eqref{eq:phaseEM} for the $m_\pi \sim 450$ MeV LQCD ensemble detailed in Ref.~\cite{Orginos:2015aya}. Shaded red and purple regions show the central value and uncertainties of $\frac{3}{2}m_\pi$ and $M_N - \frac{3}{2}m_\pi$ respectively as given in Ref.~\cite{Orginos:2015aya}. The middle plot shows the covariance of the magnitude and phase factor $\avg{C_i} - \avg{e^{R_i}}\avg{e^{i\theta_i}}$ divided by the full correlation function $\avg{C_i}$. The right plot shows the effective mass minus the sum of the magnitude and phase effective masses.}
  \label{fig:magphaseEM}
\end{figure}

LQCD calculations can be used to directly investigate the role of $\theta_i$ in the StN problem.
If $\avg{e^{i\theta_i(t)}} \sim e^{-M_N^\theta t}$ for some $M_N^\theta \neq 0$,
then the average phase then has an exponentially bad StN problem,
\begin{equation}
  \begin{split}
    \text{StN}(e^{i\theta_i}) \sim \frac{\avg{e^{i\theta_i}}}{\sqrt{\avg{|e^{i\theta_i}|^2}}} = \avg{e^{i\theta_i}} \sim e^{-M_N^\theta t}.
  \end{split}\label{eq:phaseStN}
\end{equation}
The $t$ dependence of the magnitude and phase can be studied numerically using the effective masses
\begin{equation}
  \begin{split}
    M_R(t) = -\partial_t \ln \avg{e^{R_i(t)}} ,\hspace{20pt} M_\theta(t) = - \partial_t \ln \avg{e^{i\theta_i(t)}}.
  \end{split}\label{eq:phaseEM}
\end{equation}
Fig.~\ref{fig:magphaseEM} shows results for $M_R$ and $M_\theta$
employing $N \sim 500,000$ correlation functions
previously computed by the NPLQCD collaboration
from Gaussian-smeared sources and point sinks on 
isotropic-clover gauge-field configurations 
at a pion mass of $m_\pi \sim 450~{\rm MeV}$
generated by the College of William and Mary/JLab lattice group and the NPLQCD collaboration, see Ref.~\cite{Orginos:2015aya} for further details.

At large $t$, both $M_R$ and $M_\theta$ appear approximately $t$ independent and consistent with
$M_\theta \sim M_N - \frac{3}{2}m_\pi$
and $M_R \sim \frac{3}{2}m_\pi$.
The magnitude has no significant StN problem,
while the phase has a StN problem with the same severity as the full correlation function.
As shown in Fig.~\ref{fig:magphaseEM}, 
$M - M_R - M_\theta$ is consistent with zero at large $t$. 
These results suggest the Parisi-Lepage StN problem can be identified with the StN problem arising from reweighting the complex correlation function sign problem associated with $e^{i\theta_i}$.
Similar results are expected to apply to generic complex correlation functions.
This is consistent with results for canonical and grand canonical partition functions, where reweighting approaches to sign problems lead to StN problems that become exponentially worse with increasing spacetime volume at a rate determined by an energy difference between the full theory and a phase-quenched theory~\cite{Gibbs:1986ut,Cohen:2003kd,Splittorff:2006fu,Splittorff:2007ck,Alexandru:2014hga}.

Positive-definite correlation functions is a wide variety of quantum field theories have to found to be approximately log-normally distributed.\footnote{As physical motivation, note that for weakly interacting point particles, the $n$-th moment of a correlation function describes an $n$-particle system and its energy receives one-body contributions proportional to $n$ and two-body contributions proportional to $n(n-1)/2$.
  The resulting correlation function moments, $e^{\mu n + \sigma^2 n^2 /2}$, define a log-normal distribution~\cite{Hamber:1983vu}.}
LQCD pion correlation functions are known to be approximately log-normally distributed~\cite{Hamber:1983vu} with a variance set by the $\pi\pi$ scattering length~\cite{Guagnelli:1990jb}.
Log-normal distributions have also been observed and usefully exploited in MC calculations of unitary fermions~\cite{Endres:2011jm,Endres:2011er} and condensed matter systems~\cite{Drut:2015uua,Porter:2016vry}.
At small $t$, the imaginary parts of baryon correlation functions negligible are the real parts are approximately log-normally distributed~\cite{DeGrand:2012ik}.
At large $t$, the imaginary parts are non-negligible and the real parts are better described by heavy-tailed stable distributions~\cite{davidkaplanLuschertalk}.
The distribution of the real parts can further be shown to be increasingly broad and symmetric at large $t$ by extensions of Parisi-Lepage scaling to higher moments~\cite{Savage:2010misc,Beane:2014oea}.
Similar results of log-normally distributed positive-definite correlation functions and broad, symmetric real parts of complex correlation functions are seen in QCD-like theories~\cite{Grabowska:2012ik}.

\begin{figure}[t] 
  \centering
  \includegraphics[width=\columnwidth,clip]{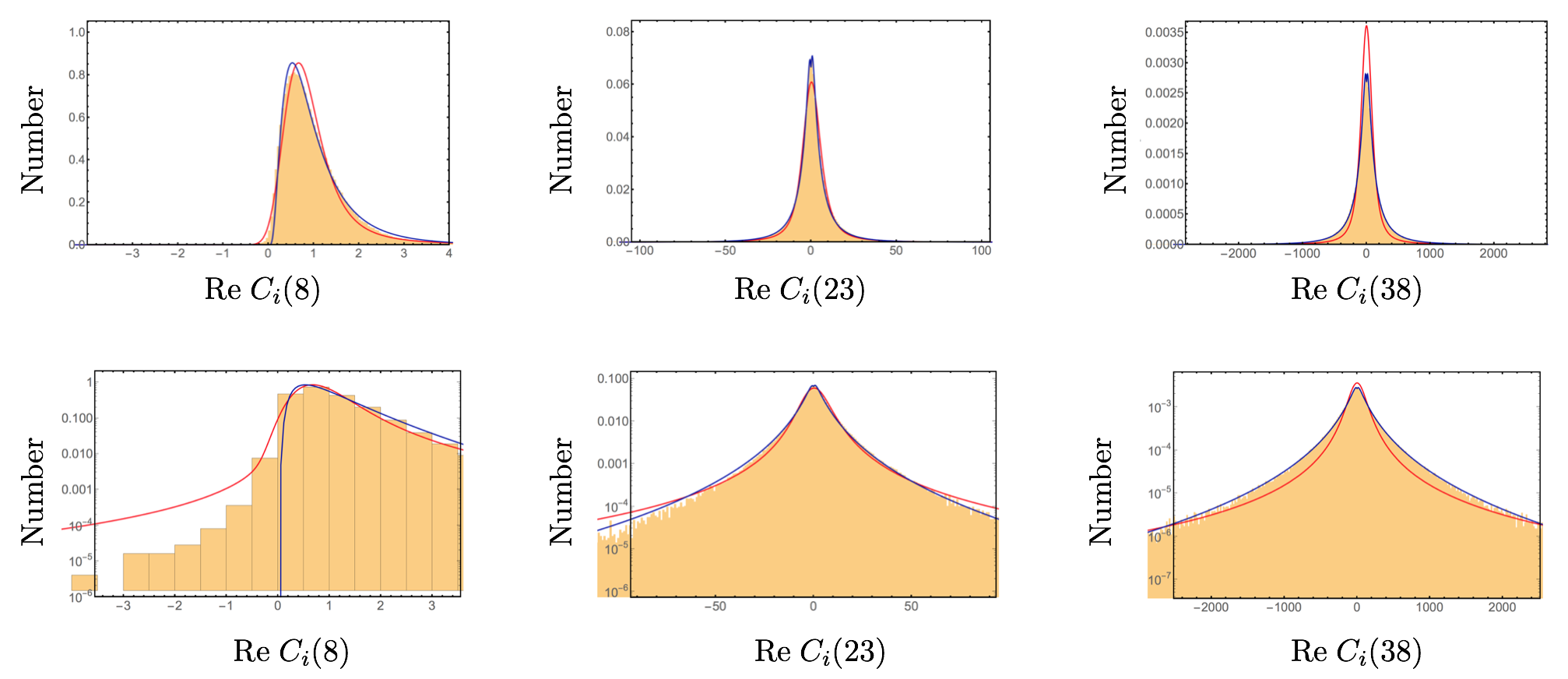}
  \caption{Histograms of the real parts of  $N=500,000$ the nucleon correlation functions at $t=8,\; 23,$ and $38$. The bottom plots only differ by having a logarithmic scale. Fits to complex-log-normal distributions with parameters determined by Eq.~\eqref{eq:CLNparams} are shown in blue. Maximum likelihood fits to stable distributions are shown in red. The vertical axis has been normalized so that the total area of the histogram is equal to unity. The horizontal axis is given in units of the average correlation function at each time and is five standard deviations wide.}
  \label{fig:rehist}
\end{figure}

LQCD calculations show that $|C_i|^2$ and further $|C_i|$ are approximately log-normally distributed at all $t$, consistent with results for positive-definite correlation functions~\cite{Wagman:2016bam}.
The phase is observed to be normally distributed at small $t$ with a variance that increases until the distribution resembles a uniform distribution over the $2\pi$ range of definition of $\theta_i$.
This suggests that complex correlation functions have a normally distributed complex log, subject to the restriction that $\theta_i$ is a circular random variable defined modulo $2\pi$.
The wrapped-normal distribution of circular statistics describes the distribution of $\theta_i$ at all $t$,
and a complex-log-normal distribution approximately describing $C_i$ can be constructed as a product 
of a log-normal magnitude and a wrapped-normal phase factor,
\begin{equation}
  \begin{split}
    \mathcal{P}(R_i,\theta_i) = \frac{1}{(2\pi)^{3/2}\sigma_R}e^{-(R_i - \mu_R)^2/(2\sigma_R^2)}\sum_{n=-\infty}^\infty e^{-n^2 (\theta_i)^2 / (2\sigma_\theta)^2},
  \end{split}\label{eq:CLN}
\end{equation}
where the parameters $\mu_R$, $\sigma_R^2$, and $\sigma_\theta^2$ are given in the infinite statistics limit by
\begin{equation}
  \begin{split}
    \mu_R &= \frac{1}{N}\sum_{i=1}^N R_i, \hspace{20pt}    \sigma_R^2 = \frac{1}{N}\sum_{i=1}^N (R_i - \mu_R)^2, \hspace{20pt} \sigma_\theta^2 = - \ln\left[ \left(\frac{1}{N}\sum_{i=1}^N \cos \theta_i \right)^2 + \left(\frac{1}{N}\sum_{i=1}^N \sin \theta_i \right)^2 \right].
  \end{split}\label{eq:CLNparams}
\end{equation}
The distribution of $\text{Re}\; C_i$ is obtained from Eq.~\eqref{eq:CLN} by performing a change of variables and marginalizing over the imaginary part.
Fig.~\ref{fig:rehist} shows histograms of the real part of baryon correlation functions along with fits to complex-log-normal distributions using Eqs.~\ref{eq:CLN}-\ref{eq:CLNparams} as well as maximum-likelihood fits to stable distributions.
Complex-log-normal distributions roughly capture the shape of the distribution at all $t$, but only describe the tails of the distribution well at large $t$.
This is expected from statistical arguments based off central limit theorems that suggest that log-normal distributions and their complex generalizations describe the large $t$ limit of correlation functions where they can be viewed as products of many independent time evolution factors~\cite{Endres:2011jm,Endres:2011er}.


In the infinite-statistics limit, parameter inference with circular random variables proceeds exactly as with ordinary random variables defined on the real line.
Away from this limit, circular statistics includes finite $N$ effects not seen for ordinary random variables.
Under the assumption of a wrapped-normal distribution, the variance of the sample mean of $\cos\theta_i$ can be computed,
\begin{equation}
  \begin{split}
    \text{StN}\left(\overline{\cos\theta_i}\right) = \frac{\avg{\frac{1}{N}\sum_{i=1}^N \cos\theta_i}}{\text{Var}\left(\frac{1}{N}\sum_{i=1}^N \cos\theta_i\right)} = \frac{e^{-\sigma_\theta^2/2}}{\frac{1}{\sqrt{2N}}\left|1 - e^{-\sigma_\theta^2}\right|} \sim \sqrt{N}\; e^{-(M_N - \frac{3}{2}m_\pi) t}.
  \end{split}\label{eq:WNcosStN}
\end{equation}
More generally, $\cos^2\theta_i = \frac{1}{2} + \frac{1}{2}\cos 2\theta_i$ and the decay of $\cos 2\theta_i = \cos(\text{arg}(C_i^2))\sim e^{-2(M_N - \frac{3}{2}m_\pi)t}$ shows that the variance of $\cos \theta_i$ includes $t$-independent contributions.
In order for this noise to not dominate the exponentially decreasing signal $\avg{\cos\theta_i}$, the statistical ensemble should be large enough that
\begin{equation}
  \begin{split}
    e^{-(M_N - \frac{3}{2}m_\pi)t} \sim \overline{\cos \theta_i} \gtrsim \frac{1}{\sqrt{N}}.
  \end{split}\label{eq:meancosbound}
\end{equation}
This bound on $\overline{\cos\theta_i}$ holds must be satisfied to avoid finite sample size effects
for generic distributions of circular random variables~\cite{Fisher:1995}.
For finite $N$ and large lattice time extent, there is a ``noise region'' where $t$ is sufficiently large that Eq.~\eqref{eq:meancosbound} is violated.
In the noise region, finite sample size effects lead to violations of Parisi-Lepage scaling and make standard estimates of ground-state energies unreliable. 
There is still useful information contained in $\theta_i$ at large $t$.
LQCD results indicate that the distribution of $\partial_t \theta_i$ approaches a $t$-independent shape at large $t$.
The wrapped normal variance of $\partial_t \theta_i$ can be computed at all $t$ and approaches a constant value at large $t$, 
but the distribution of $\partial_t \theta_i$ has heavy tails that are not adequately described by a wrapped-normal distribution.
Similar heavy tails are seen for $\partial_t R_i$.
The distributions of $\partial_t R_i$ and $\partial_t \theta_i$ can be described by heavy-tailed stable and wrapped-stable distributions respectively.
The continuum limit distributions of $\partial_t R_i$ and $\partial_t \theta_i$ are left to future work.


The finite difference interval used to define the lattice derivative can be extended beyond a single lattice spacing,
\begin{equation}
  \begin{split}
    \Delta R_i(t,\Delta t) = R_i(t) - R_i(t - \Delta t), \hspace{20pt} \Delta \theta_i(t,\Delta t) = \theta_i(t) - \theta_i(t-\Delta t).
  \end{split}\label{eq:DeltaRdef}
\end{equation}
Results for the distribution of $\Delta R_i$ with increased $\Delta t$ suggest that for $\Delta t \gg m_\pi^{-1}$,
the distribution of $\Delta R_i$ approaches a normal distribution.
This is expected for a difference of uncorrelated, normally distributed variables.
Over hadronic scales, the time evolution of $R_i$ can be described as a random walk with independent, identically distributed steps drawn from the large-$t$ distribution of $\Delta R_i$.
$\Delta \theta_i$ is fit by a heavy-tailed wrapped-stable distribution even for $\Delta t \gg m_\pi^{-1}$,
one expects that for $\Delta t \gg m_\pi^{-1}$ the time evolution of $\theta_i$
can still be treated as a random walk with independent, identically distributed steps drawn from the large-$t$ distribution of $\Delta \theta_i$.
The presence of heavy tails in $\Delta \theta_i$ signals that this random walk does not correspond to Brownian motion and is better described as a heavy-tailed L{\'e}vy flight.
The physical origin of heavy-tailed time evolution of $\ln C_i$ remains to be understood.

\section{Phase Reweighting}

Useful physical information can be extracted from the noise region if $\avg{C_i}$ can be related to properties of the $t$-independent distributions describing $\Delta R_i$ and $\Delta \theta_i$ at large $t$.
Rather than extracting the spectrum from the phase difference between the phases $\theta_i(0)$ and $\theta_i(t)$,
one could instead compare phases at very large $t$ to a more closely spaced ``source'' phase at $\theta_i(t-\Delta t)$.
By the approximate $t$-independence of the distribution of $\partial_t \theta_i$ at large $t$, results are be expected to be independent of $t-\Delta t$ provided that $\Delta t$ and $t$ are both large compared to hadronic scales.
With a source phase convention $\theta_i(0)=0$, results for $\theta_i$ can then be obtained
by taking precise results for $\Delta \theta_i$ 
in the regime $m_\pi^{-1} \ll \Delta t \ll t$ where it has a narrow, $t$-independent distribution 
and performing the extrapolation $\Delta t \rightarrow t$
to recover $\Delta \theta_i(t, t) = \theta_i(t)$.
Treating $\Delta R_i$ the same as $\Delta \theta_i$ leads to an estimator
\begin{equation}
  \begin{split}
    \tilde{M}(t, \Delta t) &= - \partial_{t + \Delta t}\ln \avg{ e^{\Delta R_i(t, \Delta t) + i\Delta \theta_i(t, \Delta t)} } = \ln \avg{\frac{ C_i(t)}{ C_i(t-\Delta t)} } - \ln \avg{\frac{ C_i(t+1)}{ C_i(t-\Delta t)} }
  \end{split}\label{eq:Mtildedef}
\end{equation}
An alternative estimator involving $\Delta \theta_i$ but instead using the full magnitude $R_i$ for any $\Delta t$ is given by~\cite{Wagman:2017xfh}
\begin{equation}
  \begin{split}
    M_{PR}(t, \Delta t) = - \partial_{t+ \Delta t}\ln \avg{ e^{R_i(t) + i\Delta \theta_i(t, \Delta t)} } = \ln \avg{C_i(t)e^{-i\theta_i(t-\Delta t)}} - \ln \avg{C_i(t+1)e^{-i\theta_i(t-\Delta t)}}.
  \end{split}\label{eq:MPRdef}
\end{equation}
This effective mass can be constructed from a phase-reweighted correlation function $G_{PR}$ as
\begin{equation}
  \begin{split}
    G_{PR}(t,\Delta t) = \avg{ C_i(t) e^{-i\theta_i(t-\Delta t)} }, \hspace{20pt} M_{PR}(t,\Delta t) = - \ln G_{PR}(t,\Delta t).
  \end{split}\label{eq:GPRdef}
\end{equation}
Physical results are recovered as $\Delta t \rightarrow t$, where $G_{PR}(t,t) = G(t)$ and $M_{PR}(t,t) = M(t)$.

\begin{figure}[t] 
  \centering
  \includegraphics[width=\columnwidth,clip]{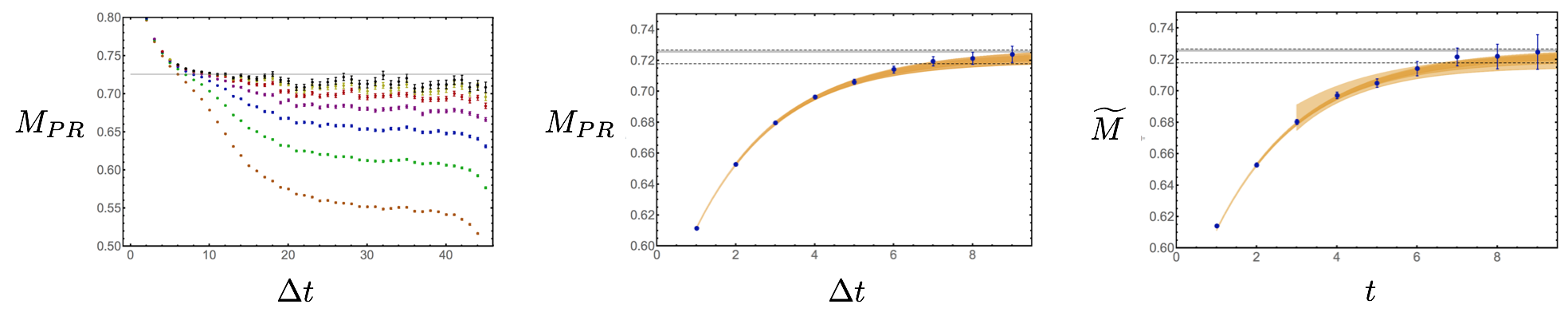}
  \caption{The left plot shows the phase reweighted nucleon effective mass $M_{PR}(t,\Delta t)$ as $t$ is varied at fixed $\Delta t$. Different colors correspond to $\Delta t = 0,\cdots,7$ in ascending order from bottom. The middle plot shows the best fit to a constant plateau of $M_{PR}$ for $t = 30 \rightarrow 40$ at various $\Delta t$ as blue points, as well as a combined fit to Eq.~\eqref{eq:MNPR} for the region $t=30\rightarrow 4$, $\Delta t = 1\rightarrow 10$ shown as a shaded region. The result for $M_N$ and its uncertainties are shown by the dashed black lines with statistical uncertainties and systematic uncertainties from variation with fitting widow range added in quadrature. The golden window result for $M_N$ and its uncertainties from Ref.~\cite{Orginos:2015aya} is shown for comparison as a gray shaded band. The right plot is analogous to the middle plot but for $\tilde{M}$ instead of $M_{PR}$. Uncertainties shown are calculated using bootstrap methods.
}
  \label{fig:PREM}
\end{figure}

Phase reweighting is similar to Green's Function Monte Carlo (GFMC) methods where the phase of the wavefunction is held fixed for initial evolution and then allowed to evolve without constraints once the system is close to it's ground state~\cite{Zhang:1995zz,Zhang:1996us,Wiringa:2000gb,Carlson:2014vla}.
Similar techniques are also used in GFMC calculations that extrapolate between a Hamiltonian without a sign problem
and the desired Hamiltonian with a sign problem~\cite{Lahde:2015ona}.
There are also similarities between phase reweighting an hierarchical integration, where locality is exploited to decompose correlation functions into products of correlation functions defined on sub-volumes that can be averaged independently~\cite{Ce:2016idq,Ce:2016ajy}.

For $\Delta t < t$, an ansatz for the parametric form of deviations between $M_{PR}$ and $M$ (analogous logic relates $\tilde{M}$ and $M$) can be deduced by splitting the correlation function time extent $[0,t]$ into a $[0, t-\Delta t]$ source construction region with unphysical time evolution where the phase is held fixed and a $[t-\Delta t, \Delta t]$ physical evolution region.
For the nucleon, one expects that
\begin{equation}
  \begin{split}
    G_{PR}(t, \Delta t) \sim e^{-\frac{3}{2}m_\pi(t-\Delta t)}e^{-M_N \Delta t}\left( 1 + \alpha \;  e^{-\delta E\; \Delta t} + \ldots  \right),
  \end{split}\label{eq:GNPR}
\end{equation}
where $\delta E$ is the gap to the first nucleon excited state that has appreciable coupling to the non-local source produced in $[0,t-\Delta t]$.
Eq.~\eqref{eq:GPRdef} turns this into an ansatz for $M_{PR}$,
\begin{equation}
  \begin{split}
    M_{PR}(t, \Delta t)  \sim M_N\left( 1 + \alpha\;  e^{-\delta E\; \Delta t} + \ldots \right).
  \end{split}\label{eq:MNPR}
\end{equation}
Fits to this form for the $m_\pi \sim 450$ MeV nucleon are shown in Fig.~\ref{fig:PREM}.
The correlation-function-ratio estimator of Eq.~\eqref{eq:Mtildedef} is also shown with fits of the same form Eq.~\eqref{eq:GNPR}.
Fits to $\tilde{M}$ give consistent results with factor of two larger uncertainties.
Phase reweighting has also been applied to two-baryon systems, and consistent results for finite-volume energy shifts have been extracted using phase reweighting and golden window methods for the $\Xi\Xi({}^1S_0)$ system.
Encouragingly, the $\Delta t$ dependence of the binding energy is much more mild than the $\Delta t$ dependence of single-particle masses shown here.

\section{Conclusion}

The StN problem is a challenge for many LQCD calculations,
particularly large nuclei made of light quarks.
LQCD calculations of the magnitudes and phases of nucleon correlation functions
show that the StN problem arises from reweighting a sign problem
and affects phases but not magnitudes.
The time evolution of the phase resembles a L{\'e}vy flight on hadronic scales, and the time derivative of the phase does not possess a StN problem.


Calculations of correlation functions reweighted with their inverse phase determined at an earlier time 
measure finite differences of phases over $t$-independent length intervals and therefore have no StN degradation with $t$.
Phase-reweighted correlation functions do have a StN problem in $\Delta t$, the length of the phase interval,
with the same severity as the standard baryon StN problem;
however, they allow additional spectral information to be extracted from correlation functions at arbitrarily large $t$.
This may prove useful in calculations of nuclei where a golden window cannot be identified.

\section*{Acknowledgements}

I thank the other members of the NPLQCD collaboration for calculations underpinning the results of this work
and for many fruitful discussions.
Calculations were performed on the Hyak High Performance Computing and Data Ecosystem at the University of Washington,
supported, in part, by the U.S. National Science Foundation Major Research Instrumentation Award, Grant Number
0922770, and by the UW Student Technology Fee (STF). 
Calculations were performed using computational resources
provided by NERSC (supported by U.S. Department of Energy grant number DE-AC02-05CH11231), and by the
USQCD collaboration. This research used resources of the Oak Ridge Leadership Computing Facility at the Oak
Ridge National Laboratory, which is supported by the Office of Science of the U.S. Department of Energy under
Contract number DE-AC05-00OR22725. The PRACE Research Infrastructure resources at the Tr\`es Grand Centre
de Calcul and Barcelona Supercomputing Center were also used. 
Parts of the calculations used the Chroma software
suite~\cite{Edwards:2004sx}. 
This work was supported in part by DOE
grant number DE-FG02-00ER41132.

\bibliography{MW_bib}


\end{document}